\documentclass[letterpaper,usenatbib]{mn2e}
\usepackage{graphicx}
\usepackage{color}
\usepackage{lscape}
\usepackage{rotating}
\usepackage{longtable}
\usepackage{epstopdf}
\newsavebox{\tablebox}
\usepackage{hyperref}
\usepackage{times}
\usepackage{longtable,lscape}
\usepackage{multirow}
\usepackage{amsmath}
\usepackage{amssymb}
\usepackage{ulem}

\newcommand{\ergs}{\ifmmode {\rm erg\ s}^{-1} \else erg s$^{-1}$\ \fi}

\newcommand{\lb}{\ifmmode L_{\rm bol} \else $L_{\rm bol}$\ \fi}
\newcommand{\ledd}{\ifmmode L_{\rm Edd} \else $L_{\rm Edd}$\ \fi}
\newcommand{\leddR}{\ifmmode L_{\rm bol}/L_{\rm Edd} \else $L_{\rm bol}/L_{\rm Edd}$\ \fi}
\newcommand{\lx}{\ifmmode L_{\rm 2-10keV} \else  $L_{\rm 2-10keV}$\ \fi}
\newcommand{\hb}{\ifmmode H\beta \else H$\beta$\ \fi}
\newcommand{\ha}{\ifmmode H\alpha \else H$\alpha$\ \fi}
\newcommand{\hg}{\ifmmode H\alpha \else H$\gamma$\ \fi}

\newcommand{\mbh}{\ifmmode M_{\rm BH}  \else $M_{\rm BH}$\ \fi}
\newcommand{\lv}{\ifmmode \lambda L_{\lambda}(1350\AA) \else $\lambda L_{\lambda}(1350\AA)$\ \fi}
\newcommand{\lcon}{\ifmmode L_{1350} \else $L_{1350}$\ \fi}

\newcommand{\mdot}{\ifmmode \dot{m} \else \dot{m} \fi }
\newcommand{\llog}{\ifmmode {\rm log} \else {\rm log} \fi }
\newcommand{\kms}{\ifmmode {\rm km\ s}^{-1} \else km s$^{-1}$\ \fi}

\begin{document}
\title[AGN and star formation activities in SDSS galaxy groups]{Effect of richness on AGN and star formation activities in SDSS galaxy groups}
\author[F. Li et al.]{Feng Li$^{1,2}$, Yi-Zhou Gu$^{1}$, Qi-Rong Yuan$^{1}$\thanks{E-mail: yuanqirong@njnu.edu.cn}, Min Bao$^{1}$, Zhi-Cheng He$^{3}$, and Wei-Hao Bian$^{1}$\\
$^1$School of Physics Science and Technology, Nanjing Normal University, Nanjing 210023, China\\
$^2$School of Mathematics and Physics, Changzhou University, Changzhou 213164, China\\
$^3$Center for Astrophysics, University of Science and Technology of China, Hefei 230026, China
} \maketitle

\begin{abstract}

Based on a large sample of 254 220 galaxies in 81 089 groups, which are selected from the spectroscopic galaxy sample of the SDSS DR12, we investigate the radial distribution of incidences, morphologies, environmental densities, and star formation properties of the active galactic nucleus (AGN) host galaxies and star-forming galaxies (SFGs) in the groups at $z<0.2$, as well as their changes with group richness ($N_{\rm rich}$).
It is found that AGN fraction slightly declines with richness for the groups/clusters. The SFG fraction is on average about 2 times larger than the AGN fraction, with a significant declining trend with richness. The group AGNs are preferentially reside in spheroidal and bulge-dominated disc galaxies, whereas the majority of SFGs are late-type discs.
Compared with the SFGs, the AGNs in poor groups ($5 \leqslant N_{\rm rich} \leqslant 10$) are closer to group center. The AGN fraction does not change with the distance to the group center, whereas the SFG fraction tends to be higher in the outskirts.
The AGNs in groups have a higher incidence than the SFGs for the massive ($\log(M_*/M_{\odot}) > 10.7$) galaxies, and the mean SFG fraction is about 6 times as that of AGNs in the late-type galaxies with lower masses at larger radius.
The distribution of environmental luminosity densities shows that the AGNs are likely to be reside in a denser environment relative to the SFGs.
Compared with the SFGs in groups, the group AGNs are found to have a higher mean stellar mass, a lower mean star formation rate, and an older mean stellar age.

\end{abstract}


\section{INTRODUCTION}

It has been appreciated that active galactic nuclei (AGNs) have played an important role in star formation and evolution of galaxies. There are many evidences that most galaxies harbor supermassive black holes (SMBHs) in their  centers \citep[e.g.][]{Koti07,Ho08,Kor-Ben09,Bluck11,Mc-Ma13}. The AGNs are powered by gas accretion onto SMBHs, but the fuel origin and the trigger mechanism of such nuclear activity is still subject to debate \citep[e.g.][]{Rees84,Shen07,Choi09}.
Several mechanisms for triggering nuclear activity in galaxies have been proposed to explain gas inflow toward centric SMBHs, such as major merger of gas-rich galaxies, minor merger of small galaxies, bar-driven gas inflow, disk instabilities, turbulence in interstellar matter, and stellar wind \citep[e.g.][]{Simk80,Bar-Her92,Elme98,Kor-Ken04,Spri05,Genz08,Hop-Qua10,Hop-Qua11}.

The distribution of AGN host galaxies in clusters, relative to the group and field environments, may provide some valuable observational constraints on AGN fueling mechanism.
Early studies showed that the AGN fraction in clusters is, on the average, smaller than that in groups or fields \citep[e.g.][]{Gisler78,Dressler85}.
Using the Sloan Digital Sky Survey (SDSS) data, \citet{Kauf04} studied the fraction of galaxies containing strong AGNs with $L_{\rm [OIII]}>10^{7}L_{\odot}$, and found that powerful AGNs are predominantly in low density regions.
Subsequent investigations also support the viewpoint that galaxies in groups retain larger reservoirs of cold gas to fuel AGN activity than their counterparts in clusters \citep[e.g.][]{Eastman07,Shen07,Georg08,Arnold09,Martini09,Allev12,Pente13,Tzana14}.
However, the conclusions about AGN fraction in clusters and fields are still controversial, and some papers show that the AGN percentage in clusters is similar to that in fields by using different AGN selection criteria \citep[e.g.][]{Haggard10,Kle-Sar12,Martini13}.

In the galaxy groups/clusters, apart from merger, harassment and ram pressure stripping are believed to be the main mechanisms that may destabilise the reservoirs of cold gas within a galaxy, and trigger the gas infall toward the central SMBH \citep{Moore96,Pogg17}.
The cluster core seems to be unfavourable to AGN \citep[e.g.][]{Gavazzi11}, which might be due to either ram pressure or temperature of intra-cluster medium (ICM) is too high to trigger an AGN \citep[e.g.][]{Davies17,Marshall18}. The galaxies that host an efficiently accreting AGN are found to be located preferentially in the infall region of projected phase-space \citep[see][for definition of infall curve]{Gordon18}, and they are rarely found in the cluster core \citep[e.g.][]{Ehlert13,Pente13,Pimbblet13,Gordon18}.
This probably implies that different gravity environments (i.e., core vs. outskirts) may have exerted considerable influence not only on possible fueling mechanism, but also on the further AGN evolution.

On the other hand, an early study performed by \citet{Per-Dys85} proposed a connection between starburst and AGN. Strong correlation between the velocity dispersion in galaxy spheroid and the SMBH mass also suggests a link between galaxy formation and the growth of central black hole \citep[e.g.][]{Magorrian98,Fer-Mer00,Gebhardt00,Tremaine02,Heckman04}. \citet{Krongold01} found that the Seyfert 2 galaxies with absence of broad (FWHM $> 1000\, \kms $) permitted lines have the same interaction frequency as star-forming galaxies (SFGs), while the Seyfert 1 galaxies with presence of broad permitted lines are in interaction less frequently. In particular, among 13 Seyfert 2 galaxies with close companions or in mergers, nine are detected by \citet{Storchi-Bergmann01} to show recent star formation in their nuclear regions. These investigations point to the scenario that the interaction is responsible for sending gas inward, which both feeds the active galactic nucleus and triggers star formation. Meanwhile, many results show that star formation and AGN activity in galaxies have roughly the same rate of evolution from $z \sim 2$ to the present \citep[e.g.][]{Boyle98,France99,Merloni04,Silver08}, indicating that AGN activity and star formation are connected, and the similar processes that trigger star formation also make a fraction of the gas available be accreted and fuel the central AGNs.

According to hierarchical scenario of the formation of large scale structures, galaxy clusters are considered to grow through the continuous accretion and merging of smaller halos across cosmic time. Member galaxies of clusters may belong to infalling groups of galaxies, i.e., a large number of members probably have been preprocessed in galaxy groups \citep[e.g.][]{Fujita04,Cortese06,Berrier09,DeLucia12,Eckert14}. Observations of the local universe suggest that natural environment of galaxies are groups and clusters. Therefore,  studies on the environmental dependence of AGN host galaxies and SFGs may shed light on evolution of galaxies and the large-scale structure of the universe \citep[e.g.][]{Tago10,Tempel12}.

Majority of galaxies in the universe are found in interacting systems, such as pairs, groups and clusters. It is very important to investigate the connection between the group environment and nuclear activity.
As hypothesized by \citet{Davies14}, the group environment is responsible for triggering gas inflow, and the environmental influence may last more than 1 Gyr. However, the resulting gas accretion on SMBHs occurs at some point during this phase, and the AGN phenomenon is on and off within a period of $\sim$ 100 Myr \citep{Shulevski15}. A random subset of active galaxies is observed in one snapshot, which might be different from another snapshot. Based on a snapshot, one might conclude that AGN fuelling is related to group environment. Since many group galaxies are inactive at any given time, the conclusion drawn from the control sample perhaps contradict with the original hypothesis. Therefore, it is necessary to explore the incidence of AGN activity as a function of group environment on the basis of a large sample of groups \citep[][]{Arnold09,Davies17}. Larger sample of the groups with various richness corresponds to a larger number of snapshots, which may reveal an global picture about the relation between AGN fuelling and group environment.

In this paper we take use of the large sample of galaxy groups, selected from the spectroscopic galaxy sample of the SDSS Data Release 12 (DR12), to investigate the projected distribution, morphologies, environmental densities, and star-formation properties (such as stellar masses, specific star formation rates, mean stellar ages) of AGN host galaxies in groups. All member galaxies in groups are spectroscopically classified into AGN host, composite, star-forming, and unclassified galaxies. Treating the group richness as a direct indicator of gravity environment on a scale of $0.1-1$ Mpc, it is interesting to see the incidences, morphologies, and above-mentioned star-formation properties for different spectroscopic classes of group galaxies as a function of the group richness. In particular, one of our concerns is to observe the difference in observational features between AGNs and SFGs within a wide scope of group richness.

This paper is structured as follows: In \S 2 we will describe the selection criteria of the group sampling and spectroscopic classification of member galaxies. The frequencies, morphologies, projected radial distributions, environmental densities, and the star-formation properties of group galaxies are presented in \S 3. Our discussion is given in \S4. The main conclusions are summarized in \S 5.  Throughout this work, we assume a flat $\Lambda$CDM cosmology with $H_0 = 67.8~\kms \rm Mpc^{-1}$, $\Omega_{\rm M} = 0.308$, and $\Omega_{\rm \Lambda} = 0.692$.

\section{Sample Selection}

\subsection{The catalog of SDSS galaxy groups}

The sample of galaxy groups we use in this paper comes from the catalog of galaxy groups achieved by \citet{Tempel17} on basis of the SDSS spectroscopic galaxy sample. The SDSS DR12 contains imaging data covering almost 14 555 $\rm deg^2$ in the $u$, $g$, $r$, $i$, and $z$ bands, and spectra for 2 401 952 galaxies and 477 161 QSOs, observed by 2.5 m telescope with a $3^\circ$ field of view at the Apache Point Observatory in Southern New Mexico. To construct a group catalog, \citet{Tempel17} use only the contiguous imaging and spectroscopic area of the Northern Galactic Cap (i.e. the Legacy Survey).

It should be noted that the spectroscopic galaxy sample is complete only up to the Petrosian magnitude $m_r$ = 17.77 \citep{Strauss02}, and incomplete for bright ($m_r < 12.5$) objects due to the CCD saturation.
After redshift correction for the motion with respect to the cosmic microwave background (CMB), the upper limit of redshift is $z = 0.2$.
Considering the missing galaxies caused by fibre collisions, \citet{Tempel17} have complemented 1 349 spectroscopic objects from previous group catalog. Based on the volume-limited sample of 584 449 galaxies, the so-called friends-of-friends (FoF) algorithm were used.  After some group membership refinements have been performed to find subgroups and expose unbound galaxies \citep[see][for details]{Tempel17}, a new catalog of groups and clusters for the SDSS DR12 has been obtained, which includes 88 662 groups with at least two members.

In general, richer groups are usually associated with more massive dark matter halos in deeper wells of gravitational potential.  As a direct observational indicator of the overdensity, the group richness ($N_{\rm rich}$) may represent the gravity environment on a scale of $0.1-1$ Mpc.
However, the richness separatrix for group/cluster separation is somewhat ambiguous in the literature. \citet{Abell58} early defined a cluster which contains at least 50 member galaxies within the magnitude range between $m_{3}$ and $m_{3}+2$, where $m_{3}$ is the magnitude of the third brightest member galaxies.
Then, \citet{Abell89} defined a cluster with more than 30 galaxies in ($m_{3}$, $m_{3}+2$) range. Theoretically,
\citet{Bower-Balogh04} defined galaxy group/cluster based on the mean virialized halo mass: a cluster means the halo mass greater than $10^{14} M_{\odot}$, and a group with halo mass in the range from $10^{13} M_{\odot}$ to $10^{14} M_{\odot}$. \citet{Eke06} defined a cluster with a higher separatrix of virial mass, $\sim 10^{14} h^{-1} M_{\odot}$.
Recently, the X-ray luminosity at 0.1-2.4 keV is taken to derive the group mass on the basis of the empirical mass-luminosity relation \citep{Leauthaud10}. With this $L_{\rm X}-M$ relation, \citet{Allev12} derived the masses of 189 X-ray-detected galaxy groups with $41.3 < \log (L_{\rm X}/{\rm erg}~{\rm s}^{-1}) < 44.1$ at 0.1-2.4 keV  in the COSMOS field, and found that all these groups have their halo masses in a range of $13 < \log (M_{200}/M_{\odot}) < 14.5$.

\begin{figure}
\begin{center}
\includegraphics[width=8cm]{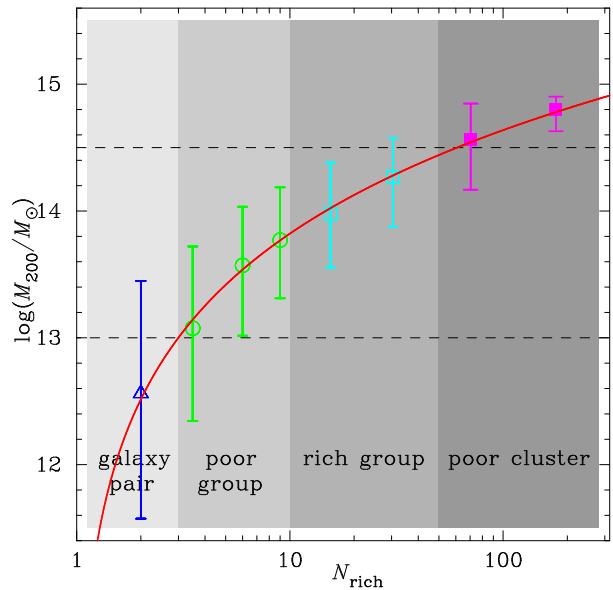}
\caption{The median of group mass, ${\rm log}(M_{200}/M_{\odot})$, as functions of group richness, $N_{\rm rich}$. Blue triangle, green circles, cyan rectangles, and pink solid rectangles denote galaxy pairs, poor groups, rich groups and poor clusters, respectively. Error bars are the median values of the deviation from median mass in each $N_{\rm rich}$ bin. The region between two dashed lines corresponds to the domain of galaxy groups with $13 < \log (M_{200}/M_{\odot}) < 14.5$.}
\label{fig10}
\end{center}
\end{figure}

The virial mass within virial radius in the SDSS group sample have been estimated by \citet{Tempel17}. To quantitatively define a variety of gravity environments with richness, we firstly show the relation between richness ($N_{\rm rich}$) and virial mass ($M_{200}$) in Fig.~\ref{fig10}. Our sample of groups is divided into eight subsamples according to group richness (see Table \ref{table1}).
Fig.~\ref{fig10} gives the median of group mass, ${\rm log}(M_{200}/M_{\odot})$, as a function of group richness, $N_{\rm rich}$. The galaxy groups defined by $13 < \log (M_{200}/M_{\odot}) < 14.5$ are found to have richness ($N_{\rm rich}$) ranging from 3 to 50.
Therefore, we refer to the galaxy overdensity systems with $3 \leq N_{\rm rich} \leq 50$ as groups, and those with $ N_{\rm rich} > 50$ are just termed as poor clusters. Empirically, we refer to the groups with $3 \leqslant N_{\rm rich} \leqslant 10$ as poor groups, and refer to the groups with $11 \leqslant N_{\rm rich} \leqslant 50$ as rich groups (see Fig.~\ref{fig10}).

\subsection{Classification and physical parameters of member galaxies}

By providing a homogeneous spectroscopic data of high quality, the SDSS allows statistical studies of the physical properties of galaxies and AGNs on the basis of objective spectral analysis.  Estimation of physical parameters for the SDSS DR7 galaxies in above-mentioned contiguous regions of the Legacy Survey has been listed in the MPA/JHU spectroscopic catalog $^{}$\footnote{$^{}$ The catalog of MPA/JHU physical parameters is available at \url{http://wwwmpa.mpa-garching.mpg.de/SDSS/DR7/}}. The information about spectral classifications based on the standard emission line ratio diagnostic diagrams, the Baldwin-Phillips-Terlevich (BPT) diagrams \citep{BPT81} is given. Additionally, they provide some important star formation properties, such as stellar masses ($M_*$), star formation rates (SFRs), the strength of the continuum break at 4000 \AA~ in galaxy spectra ($D_{n}$, as an indicator of average stellar age) \citep{Kauf03a,Brin04}. All these galaxy properties have been corrected for AGN emission (see \S \ref{prop} below). Although the MPA/JHU spectroscopic catalog has not been updated to the DR12, more than 96\% member galaxies in \citet{Tempel17} group catalog are included in the MPA/JHU database.

According to the unified model of AGN, AGNs can be separated into two categories. For the Type 1 AGNs,  central black holes and the associated continua are viewed directly, as well as the broad-line regions. Only the narrow-line regions can be observed for the Type 2 AGNs due to their obscuring media. It is a challenge to study the  stellar population in host galaxies of type 1 AGNs,  so  \citet{Kauf03c} excluded the type 1 AGNs in the MPA/JHU catalog. The BPT line-ratio diagram allows us to distinguish Type 2 AGNs from normal SFGs by considering the intensity ratios of two pairs of relatively strong emission lines.
Based on the BPT diagrams developed by \citet{Kauf03c} and \citet{Kew01}, the SDSS galaxies with signal-to-noise S/N $\geqslant 2$ in H$\alpha$ are spectroscopically classified into following six categories: SFG, low S/N SFG, composite, AGN, low S/N AGN, and `UnClass'.
In this paper, regardless of whether the signal-to-noise is high or low, all group galaxies are classified as four populations: SFG, composite, AGN, and UnClass. `UnClass' population represents the galaxies with no or very weak emission lines, and it is impossible to classify them with the BPT diagram. Our subsequential analysis will show that these unclassified galaxies are predominated by the early-type galaxies (i.e. E and S0 galaxies) (see \S 3.2). After a cross-identification of the group galaxies with the MPA/JHU catalog, we find the majority ($\sim 96.5\%$) galaxies in SDSS DR12 group catalog have their counterparts in the MPA/JPU data. We construct our group sample by including the groups in which more than 90\% member galaxies are listed in the MPA/JPU data. As a result, there are 81 089 groups and 254 220 member galaxies in our sample, including about $91.5\%$ of groups in \citet{Tempel17} catalog. Table \ref{table1} presents the statistics of groups/clusters and member galaxies as a function richness.  Among all member galaxies in our sample, 110 645 ($43.5\%$) galaxies are classified as the SFGs,
35 339 (13.9$\%$) galaxies are found to host AGNs. 16 033 (6.3$\%$) and 92 203 (36.3$\%$) galaxies are classified as composite and `UnClass' populations, respectively.

\section{Properties of group AGNs and SFGs}

\subsection{Effect of richness on AGN and SFG fractions}

To explore the environmental effect on AGN incidence in the groups with a wide range of richness, our group sample is divided into eight subsamples according to group richness. The fractions of four classes of galaxies  (i.e. AGN, Composite, SFG, and `UnClass') can be calculated for different richness bins. Assuming that the galaxy counts is a Poisson variable \citep[]{Silver08}, the error bar of fraction can be estimated by error-propagation.
The fractions of four classes of galaxies ($f_{\rm SFG}$, $f_{\rm Comp}$, $f_{\rm AGN}$ and $f_{\rm UnClass}$) for different richness bins are listed in Table \ref{table1}. Additionally, in order to facilitate the comparison with the statistics of field galaxies, we construct a sample of 293 887 field galaxies on basis of our cross-identification of the galaxies in \citet{Tempel17} catalog with the MPA/JHU database. Their fractions of four spectroscopic subclasses are also given in Table \ref{table1}.

\begin{figure}
\begin{center}
\includegraphics[width=7cm]{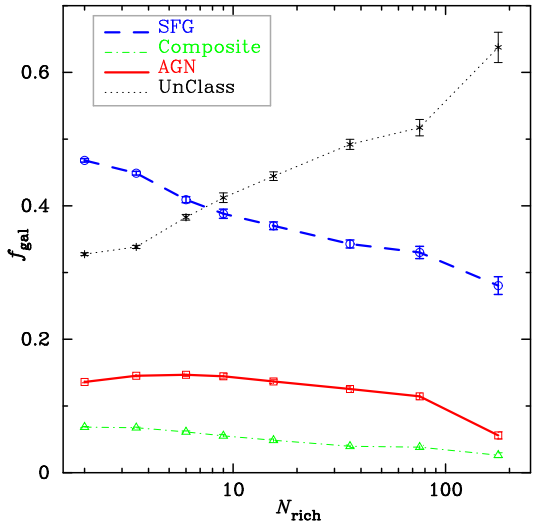}
\caption{The fractions of different classes of galaxies, $f_{\rm gal}$, as functions of group richness, $N_{\rm rich}$. Blue circles (dashed line), green triangles (dot-dashed line), red rectangles (solid line), and black crosses (dotted line) denote the SFGs, composites, AGNs and `UnClass' galaxies, respectively. }
\label{fig1}
\end{center}
\end{figure}

\begin{center}
\begin{table*}
\centering \caption{Sizes and fractions of different galaxy classes for eight richness bins. }
\label{table1}
\begin{tabular}{|c|c|c|c|c|c|c|c|}\hline\hline

groups/clusters & Richness & $N_{\rm group}$ & $N_{\rm galaxy}$ & $f_{\rm SFG}$ & $f_{\rm Comp}$ & $f_{\rm AGN}$ & $f_{\rm UnClass}$ \\ \hline
field galaxy & --- & --- & 293 887 & $ 49.78(\pm0.16)$\% & $6.49(\pm0.05)$\% &$12.37(\pm0.04)$\% & $31.36(\pm0.12)$\% \\ \hline
galaxy pair &     $2$ & 48 720 & 97 440 & $46.81(\pm0.27)$\% & $6.83(\pm0.09)$\% &$13.60(\pm0.13)$\% & $32.76(\pm0.21)$\% \\ \hline
\multirow{3}{*}{poor group}
 & $[3,4]$ & 24 324 & 81 371 & $44.89(\pm0.28)$\% & $6.74(\pm0.09)$\% &$14.53(\pm0.14)$\% & $33.84(\pm0.24)$\% \\
 & $[5,7]$ &  5 089 & 28 897 & $40.92(\pm0.45)$\% & $6.12(\pm0.15)$\% &$14.68(\pm0.24)$\% & $38.28(\pm0.43)$\% \\
 & $[8,10]$ &  1286 & 11 248 & $38.82(\pm0.69)$\% & $5.52(\pm0.23)$\% &$14.45(\pm0.38)$\% & $41.21(\pm0.72)$\% \\ \hline
\multirow{2}{*}{rich group}
 & $[11,20]$ &  1 129 & 15 256 & $37.01(\pm0.58)$\% & $4.86(\pm0.18)$\% &$13.68(\pm0.32)$\% & $44.45(\pm0.65)$\% \\
 & $[21,50]$ &   446 & 12 788 & $34.28(\pm0.60)$\% & $3.96(\pm0.18)$\% &$12.54(\pm0.33)$\% & $49.21(\pm0.76)$\% \\ \hline
\multirow{2}{*}{poor cluster}
 & $[51,100]$ &    80 &  5 198 & $33.01(\pm0.92)$\% & $3.83(\pm0.28)$\% &$11.45(\pm0.49)$\% & $51.71(\pm1.23)$\% \\
 & $[101,254]$ &    15 &  2022 & $28.04(\pm1.33)$\% & $2.62(\pm0.37)$\% &$ 5.59(\pm0.54)$\% & $63.75(\pm2.27)$\% \\ \hline
total sample & [2,254]  & 81 089 & 254 220  & $43.52(\pm0.16)$\%  & $6.31(\pm0.05)$\%  & $13.90(\pm0.08)$\%  & $36.27(\pm0.14)$\% \\ \hline

\end{tabular}
\end{table*}
\end{center}

Fig. \ref{fig1} gives the trends of AGN and SFG fractions along with group richness. The AGN fraction is $13.6\%$ for galaxy pairs ($N_{\rm rich}=2$), similar to those of poor groups with $3 \leqslant N_{\rm rich} \leqslant 10$. For the large sample of field galaxies, the AGN fraction is about 12.37\% (see Table \ref{table1}), slightly lower than those for galaxy pairs.
No significant excess in $f_{\rm AGN}$ is found for galaxy pairs, which is consistent with many previous studies \citep[e.g.][]{Schmitt01,Kelm04,Grogin05,Col-Lam06,Li08}.
However, some investigations show that the AGN incidence is enhanced in galaxies in close pairs \citep[e.g.][]{Storchi-Bergmann01,Koul06,Alonso07,Urru08,Comer09,Rogers09,Koss10,Ellison11,Silver11,Sabater15}. For example, \citet{Alonso07} find a higher fraction of Type 2 AGN in close galaxy pairs (30\%), relative to the isolated sample (23\%), suggesting a strong connection between galaxy interaction and nuclear activity.
Since we do not have the information on the state of the interaction in the pairs, our result does not mean that close interaction between galaxies is not favourable at triggering AGNs.

As shown in Fig. \ref{fig1}, $f_{\rm AGN}$ seems to be constant, $\sim 14.56 (\pm 0.12)\%$, for poor groups with $3 \leqslant N_{\rm rich} \leqslant 10$, which implies that the AGN incidence of in low-$z$ poor groups is insensitive to the gravity environment. However, for all groups and poor clusters with $3 \leqslant N_{\rm rich} \leqslant 100$, a very slight decline of $f_{\rm AGN}$ can be linearly fitted as $f_{\rm AGN} = (-0.025 \pm 0.004)\, \log N_{\rm rich} + (0.165 \pm 0.004)$, and the Pearson correlation coefficient is $-0.962$. We studied the fraction of X-ray selected AGNs with optical emission lines for the X-ray groups, and found a clear decreasing trend with group richness at $z < 0.4$ in the COSMOS field \citep{Li17}. The discrepancy in declining slope is likely to be caused by different redshift range and selection criteria of groups and AGNs.
It should be noticed that there is a significant drop (by a factor of 2) in AGN fraction ($5.59\%$) at $N_{\rm rich} > 100$, indicating a significant difference in AGN incidence in cluster environment. The cluster environment tends to be hostile to host AGNs, supporting the point that the galaxies in groups seem to be apt to retain larger reservoirs of cold gas to fuel AGN activity than the galaxies in clusters \citep[e.g.][]{Shen07,Georg08,Arnold09,Martini09,Allev12,Tzana14}.

In comparison, the SFG fraction decreases significantly with group richness, and the linear fitting is $f_{\rm SFG} = (-0.270 \pm 0.028) {\rm log}N_{\rm rich} + (0.568 \pm 0.020)$, with a Pearson correlation coefficient of $-0.971$. On the average, the number of SFGs is about 2 times larger than that of AGNs. The SFGs are found to be predominant in the poor groups with $3 \leqslant N_{\rm rich} \leqslant 10 $ in the local universe.
The fraction of composite galaxies shows a slight decreasing trend, from 6.8\% to 2.6\%, as the group richness increases.
For richer groups, the `UnClass' galaxies with a greater percentage are found to be dominated by early-type galaxies (see section3.2). This trend is consistent with previous findings that the star formation activity depends on environmental density, and the quenching of star formation tends to be more efficient in high density regions \citep{Cucci06,Peng10,Scovi13,Tal14,Gu18}. Harassment and ram pressure stripping have been proposed to be the main mechanisms to suppress star formation activity for a galaxy in rich groups/clusters \citep{Moore96,Pogg17}.

\subsection{Morphological classification of group galaxies}

The connection between AGN phenomenon and morphology of the host galaxies can be of great help in understanding the origin of the fuel and the trigger mechanisms of AGN activity. \citet[][hereafter HC11]{HC11} developed a classification method based on a machine learning algorithm, support vector machines (SVM)\citep{HC08,HC09}, which is proved to be better adapted to nature after being compared to several independent visual classifications, such as \citet{Nair-Abraham10} catalog of $\sim 14~000$ SDSS galaxies with $m_{g} < 16$, and the {\rm Galaxy Zoo} first release catalog by \citet{Lintott11}.
According to the probabilities of being E, S0, Sab, Scd galaxies (i.e. $P_{\rm E}$, $P_{\rm S0}$, $P_{\rm Sab}$, and $P_{\rm Scd}$) that were computed by HC11, all group galaxies are classified into four subclasses (i.e. E, S0, Sab, and Scd).

\begin{figure}
\begin{center}
\includegraphics[width=8cm]{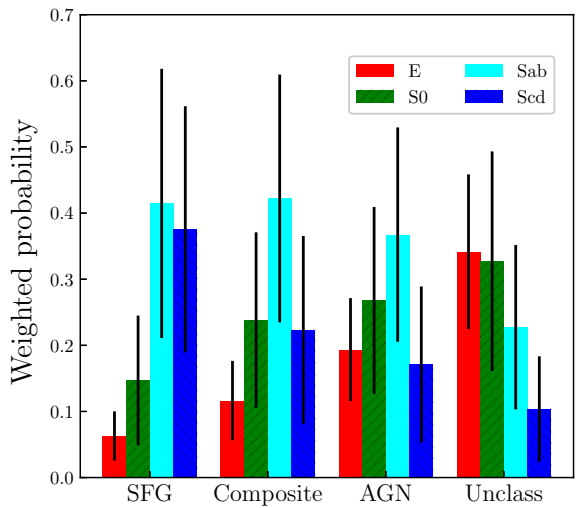}
\caption{The weighted probability distribution of different morphology subclasses for four spectroscopic classes of group galaxies. The E, S0, Sab and Scd galaxies are denoted in red, green, cyan and blue colors, respectively.}
\label{fig2}
\end{center}
\end{figure}

The morphological subclass with weighted probability is assigned to each group galaxy. Fig. \ref{fig2} shows the normalized distributions of weighted probability being each morphological type for four spectroscopic classes. For a specified morphology subclasses, the weighted probability and its 1-$\sigma$ error can be computed according to Equations (5) and (6) in HC11, which are listed in Table \ref{table2}.

\begin{center}
\begin{table*}
\centering \caption{Normalized weighted probabilities of morphology for different spectroscopic classes of galaxies}
\label{table2}
\begin{tabular}{c|c|c|c|c|c}\hline\hline

\multirow{2}{*}{Galaxy Class} & \multirow{2}{*}{Number } & \multicolumn{4}{|c|}{Normalized Weighted Probabilities } \\ \cline{3-6}
         &         &   E   &   S0  &  Sab  &  Scd  \\ \hline
   SFG  & 110645  & $6.29(\pm3.72)$\%  & $14.69(\pm9.80)$\%  & $41.46(\pm20.36)$\% & $37.56(\pm18.60)$\% \\
Composite & 16033 & $11.64(\pm6.00)$\% & $23.82(\pm13.27)$\% & $42.21(\pm18.74)$\% & $22.33(\pm14.21)$\% \\
   AGN  & 35339  & $19.36(\pm7.78)$\%  & $26.80(\pm14.13)$\% & $36.74(\pm16.21)$\% & $17.10(\pm11.80)$\% \\
UnClass & 92203 & $34.17(\pm11.68)$\% & $32.71(\pm16.61)$\% & $22.74(\pm12.43)$\% & $10.38(\pm7.96)$\% \\ \hline
\end{tabular}
\end{table*}
\end{center}

It should be noticed that the behavior of the learning algorithm depends on how close the training sample is to the real sample that one wants to classify. HC11 took the SDSS galaxies with visual classifications by \citet[][hereafter F07]{Fukugita07} as the training sample.
To verify how good this learning algorithm at classifying the Type 2 AGNs in our sample, we perform a cross-identification of this training sample of galaxies in F07 with the MPA/JHU catalog, and 2 227 ($\sim 98.8\%$) galaxies are found to have counterparts. There are 503 ($\sim 22.6\%$) AGN host galaxies in the training sample on the basis of the BPT diagnostic diagrams.

It is important to define an independent test set of AGN hosts with known morphological classifications to measure how well the SVM method by HC11 works in morphological classification of the AGNs in our sample.
Following HC11, we train the SVM algorithm with a sample of 503 AGN hosts with visual classifications by F07, and test the robustness and accuracy of the trained SVM model with a testing AGN sample given by \citet{Nair-Abraham10}. Then, we compare the probability distributions of the AGN testing sample that are produced by the SVM algorithms trained independently with above two samples (i.e., AGN sample vs. sample of all galaxies). It is found that the effect of changes in the training set on final classification is very slight, demonstrating that our SVM algorithm trained with the SDSS galaxies in F07 is competent for morphological classification of Type 2 AGN host galaxies.

From Fig. \ref{fig2} and Table ~\ref{table2}, it can be seen that early-type galaxies  (E $+$ S0) are predominated in `UnClass' population, and the composite galaxies are dominated by early disc galaxies (Sab), followed by spheroidal and late disc galaxies (S0 $+$ Scd).
For the SFGs in groups, the late-type disc galaxies (Sab $+$ Scd) are predominated, whereas the total weighted probabilities of early-type galaxies appear very small. The AGN host galaxies seem to have a very high probability of being the spheroidal and early disc galaxies (S0 $+$ Sab) with notable bulges, and have  small probability to reside in elliptical and late-type disc galaxies (E $+$ Scd).
Our results are consistent with the previous studies based on the SDSS data. \citet{Kauf03c} find that the AGNs at $0.02 < z < 0.3$ are predominantly hosted in the SDSS galaxies with massive bulges. Based on the SDSS DR5 data, most AGNs at $0.025 < z < 0.107$ are found to reside in late-type galaxies with intermediate luminosities and velocity dispersions \citep[][]{Choi09}.

Our statistics is also in accordance with the results based on the sample of X-ray selected AGNs.
By using HST/ACS images and a photometric catalog in the COSMOS field at $0.3 < z < 1.0$, \citet{Gabor09} studied the S\'ersic index distributions for the host galaxies of X-ray selected AGNs, finding that approximately 60\% of all the control samples have late-type morphologies, and the remaining 40\% have early-type morphologies.
With the public Chandra X-ray data, \citet{Pierce07} found that $53^{+11}_{-10}\%$ of the X-ray selected AGNs reside in E/S0/Sa galaxies, and \citet{Povic09} drew a conclusion that the AGNs in early-type galaxies have lower Eddington rates than those in late-type galaxies. Using a sample of 262 AGNs in the Subaru/XMM-Newton Deep Survey (SXDS), \citet{Povic12} found that at least 50\% of X-ray detected AGNs at $z \leqslant 2.0$ reside in spheroidal and bulge-dominated galaxies, while at least 18\% are hosted in disc-dominated galaxies, suggesting that different mechanisms may be responsible for triggering the nuclear activity.

\subsection{Radial distribution of the AGNs and SFGs in groups}

To analyze the distribution of different classes of galaxies in groups, the projected radius of galaxies, $R$, from group center should be calculated. The projected radius can be used to trace environmental density.  As a proxy of the virial radius, $R_{200}$ is originally defined to study the properties of galaxy clusters \citep{Yan15}. By definition, $R_{200}$, is the radius inside which the cluster mass density is 200 times the critical density of the universe, $\rho_{c}(z)$ \citep[][]{Finn05}:
\begin{equation}
200\rho_{c}=\frac{M_{\rm cl}}{\frac{4}{3}\pi R_{200}^{3}},
\end{equation}
where $M_{\rm cl}$ is the total mass of the cluster. \citet{Finn05} used the redshift dependence of the critical density and the virial mass to relate the line-of-sight velocity dispersion of the cluster, $\sigma_{r}$, to the cluster mass, so that $R_{200}$ becomes \citep[also see][]{Carlberg97}:
\begin{equation}
R_{200}=\frac{\sqrt{3} \sigma_{r}}{10H_{0}\sqrt{\Omega_{\rm \Lambda}+\Omega_{\rm M}(1+z_{\rm cl})^{3}}} \quad {\rm Mpc}
\end{equation}

For galaxy groups, the virial radii $R_{\rm vir}$ of the groups are calculated by following formula \citep{Tempel14}:
\begin{equation}
\frac{1}{R_{\rm vir}}=\frac{2}{(1+z_{\rm gr})n(n-1)}\sum_{i \neq j}^{n}\frac{1}{R_{ij}},
\end{equation}
where $R_{ij}$ is the projected distance between galaxies in pairs in a group, $z_{\rm gr}$ is the mean group redshift. For the compact groups with $2 \leqslant N_{\rm rich} \leqslant 4$, the virialization may have not been completed, thus the uncertainty of virial radius is expected to be greater. \citet{Tempel17} regard this virial radius as $R_{200}$ \citep[see][for details]{Tempel14}.

\begin{figure}
\center
\includegraphics[width=9cm]{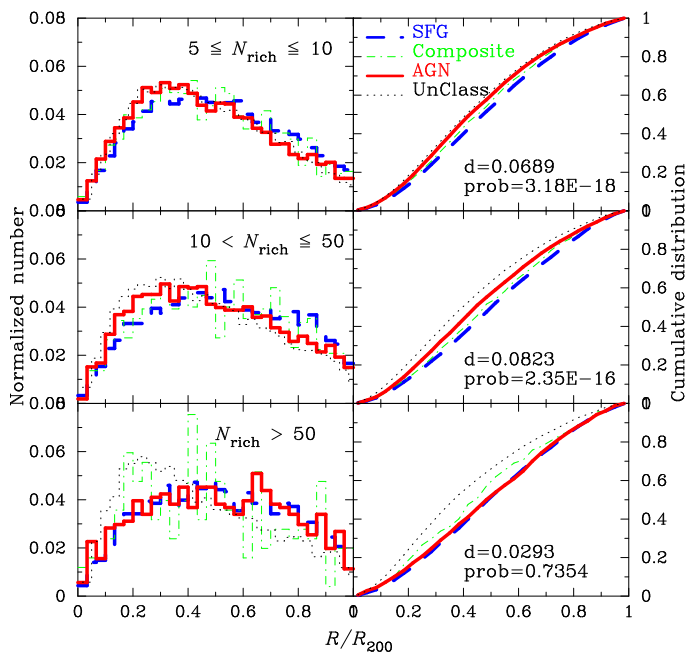}
\caption{The normalized number of all classes of galaxies in poor groups ($5 \leqslant N_{\rm rich} \leqslant 10$), rich groups ($10 < N_{\rm rich} \leqslant 50$), and poor clusters ($N_{\rm rich} > 50$), as a function of the normalized group-centric radius. The probabilities and the $d$ values in the Kolmogorov-Smirnov test are given on right panels. Blue dashed lines, green dot-dashed lines, red solid lines, and black dotted lines denote the SFGs, composites, AGNs, and `UnClass' galaxies, respectively.}
\label{fig3}
\end{figure}

For each group galaxy, we calculate the projected group-centric radius, $R$, and then normalize to $R_{200}$. We take $R/R_{200}$ as normalized radial location within a group. Fig. \ref{fig3} gives the distribution of the normalized group-centric radius, $R/R_{200}$, for four classes of galaxies in the groups with different richness. To avoid the bias in $R_{200}$ statistics for poor groups with $2 \leqslant N_{\rm rich} \leqslant 4$, we adopt three richness bins: $5 \leqslant N_{\rm rich} \leqslant 10$, $10 < N_{\rm rich} \leqslant 50$, and $N_{\rm rich} > 50$, representing poor groups, rich groups, and poor clusters, respectively. It is found that the AGNs prefer to locate at $0.2\sim0.4 R_{200}$ in poor groups with $5 \leqslant N_{\rm rich} \leqslant 10$, with a peak of normalized distribution function at $\sim 0.3 R_{200}$, which is quite similar with the situation of early-type `UnClass' population. The radius distribution of group SFGs peaks at a larger radius for the same richness bin ($5 \leqslant N_{\rm rich} \leqslant 10$), showing a different distribution of projected radius.
As the richness increases, we can see that the AGN distribution is slightly away from the center, but still within $0.3\sim0.7 R_{200}$.
The Kolmogorov-Smirnov (K-S) tests are performed for the SFGs and AGNs in different richness ranges, and find that the SFGs and AGNs in poor and rich groups ($5 \leqslant N_{\rm rich} \leqslant 50$) have different distributions of the group-centric radius, with small K-S probabilities.
For the poor clusters with $N_{\rm rich} > 50$, the early-type `UnClass' galaxies tend to be more concentrated into the group centers, and a high K-S probability ($73.5\%$) for member SFGs and AGNs shows that they are very unlikely to have different radius distributions in poor clusters. In comparison to the groups, there are more AGNs in the outskirts of poor clusters.

Does it imply that the AGNs in clusters are likely to be triggered in more peripheral regions? Compared with the group environment, however, the AGNs in clusters exhibit a completely distinct projected distribution. \citet{Martini07} found that most luminous AGNs ($L_{\rm X} > 10^{42}$ \ergs) in clusters are more centrally concentrated than other cluster galaxies with similar luminosities, which seems to be contrary to the expectation that most AGNs will be triggered in the outskirts of clusters. For the high-$z$ galaxy clusters ($z > 0.5$), \citet{Galametz09} found an AGN overdensity at $R < 0.5$ Mpc, by using the data of the NOAO Deep Wide-Field Survey. It has been proposed that feedback from the central AGNs in clusters may drive bulk motion pushing the central gas farther outwards, which will lead to higher number of the SFGs in outer regions \citep{Wang10,Tholken16}.

\begin{figure}
\center
\includegraphics[width=9cm]{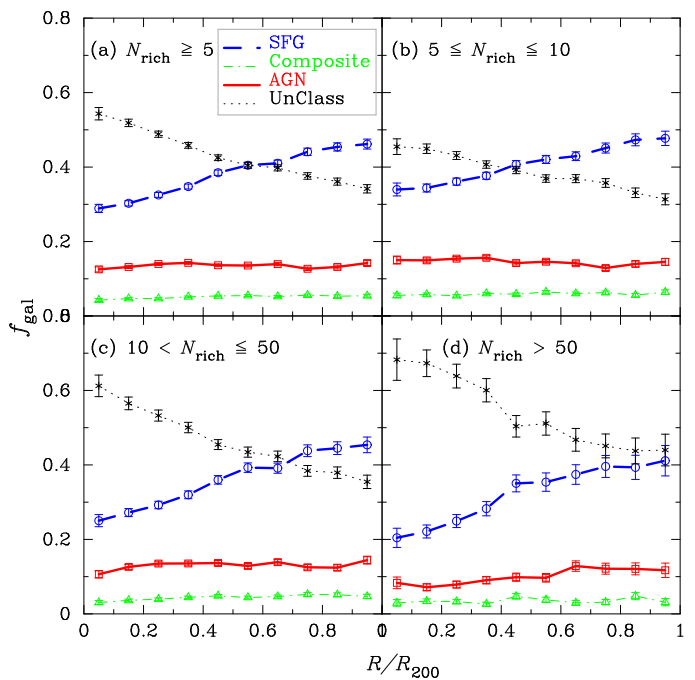}
\caption{The fractions of different classes of galaxies, $f_{\rm gal}$, as a function of normalized radius, $R/R_{200}$, for the groups/clusters with $N_{\rm rich} \geqslant 5$. Panel (a) present the overall radial distribution of galaxy fractions, and panels (b),(c), and (d) present those for poor groups ($5 \leqslant N_{\rm rich} \leqslant 10$), rich groups ($10 < N_{\rm rich} \leqslant 50$), and poor clusters ($N_{\rm rich} > 50$), repsectively. Blue circles (dashed line), green triangles (dot-dashed line), red rectangles (solid line), and black crosses (dotted line) denote the SFGs, composites, AGNs, and `UnClass' galaxies, respectively. }
\label{fig4}
\end{figure}


\begin{figure}
\center
\includegraphics[width=7cm]{fig5-2.eps}
\caption{The fractions of different classes of (a) less massive ($\log(M_*/M_{\odot}) \leqslant 10.7$) and (b) massive ($\log(M_*/M_{\odot})>10.7$) galaxies as a function of normalized radius, $R/R_{200}$, for the groups/clusters with $N_{\rm rich} \geqslant 5$. Blue circles (dashed line), green triangles (dot-dashed line), red rectangles (solid line), and black crosses (dotted line) denote the SFGs, composites, AGNs, and `UnClass' galaxies, respectively. }
\label{fig5}
\end{figure}


\begin{figure}
\center
\includegraphics[width=7cm]{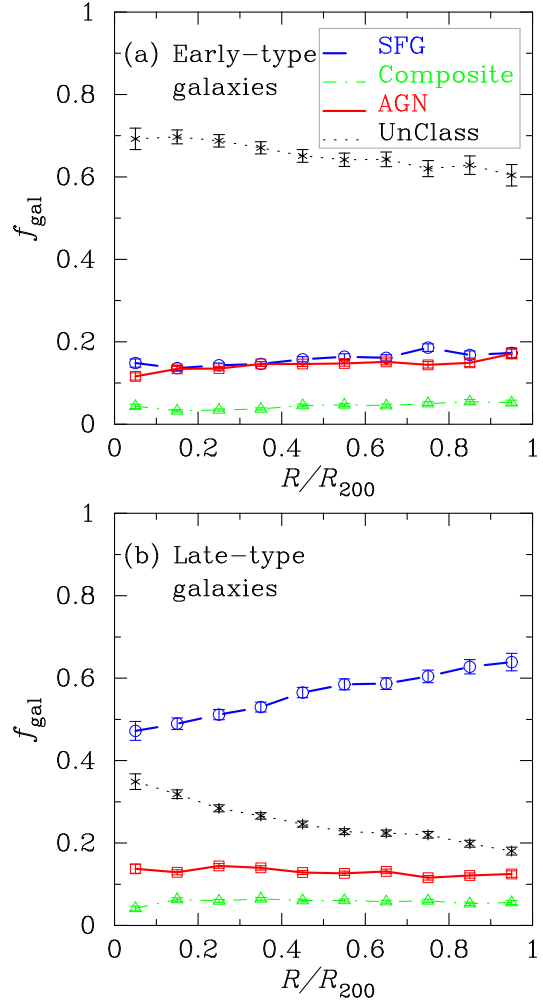}
\caption{The fractions of different classes of (a) early-type and (b) late-type galaxies, $f_{\rm gal}$, as a function of normalized radius, $R/R_{200}$, for the groups/clusters with $N_{\rm rich} \geqslant 5$. Blue circles (dashed line), green triangles (dot-dashed line), red rectangles (solid line), and black crosses (dotted line) denote the SFGs, Composites, AGNs, and `UnClass' galaxies, respectively. }
\label{fig6}
\end{figure}

Fig. \ref{fig4} presents the fractions of different classes of galaxies in the groups/clusters with $N_{\rm rich} \geqslant 5$, as a function of the group-centric radius normalized to the group virial radius, $R/R_{200}$. The overall radial profiles of population fraction are given in the panel (a). The SFG fraction tends to increase with the projected radius, and a contrary trend is found for early-type galaxies in `UnClass' population. On the other hand, it shows that AGN fraction does not vary with the normalized projected radius, with a constant overall AGN fraction of $\sim 13\%$. Composite galaxies are found to have a constant fraction of $\sim 5\%$.
To observe the dependence upon richness, we split our sample into three $N_{\rm rich}$ bins (say [5,10],(10,50], and $>$50), representing poor groups, rich groups, and poor clusters, repsectively. Their fraction profiles are presented in the panels (b), (c), and (d), respectively. As the group richness increases, the SFG fraction tends to be lower, and the $f_{\rm SFG}$ difference between the inner ($R<0.4 R_{200}$) and outer ($R \geqslant 0.4 R_{200}$) regions tends to be more significant. No significant changes in the fractions of AGNs and composite galaxies are found for the groups with  $5 \leqslant N_{\rm rich} \leqslant 50$. For the poor clusters with $N_{\rm rich} > 50$, a lower AGN fraction is perceptible in the inner regions ($R < 0.6 R_{200}$).

For typical clusters of galaxies, previous investigations show different radial distributions of AGN fraction for various cluster samples. Based on 6 self-similar SDSS galaxy clusters at $z \sim 0.07$, \citet{Pimbblet13} found that AGN fraction increases significantly from the cluster centre to $1.5 R_{\rm vir}$, but flattens off quickly at larger radii. Additionally, they studied a mass dependence of radial distribution of AGN fraction, and massive ($\log (M_*/M_{\odot}) > 10.7$) galaxies are found to have systematically higher AGN fractions than lower mass galaxies at all radii from cluster center.
To verify the mass dependence of AGN fraction in the SDSS groups, we simply split our sample of member galaxies in groups/clusters ($N_{\rm rich} \geqslant 5$) into the low-mass ($\log(M_*/M_{\odot}) \leqslant 10.7$) and high-mass ($\log(M_*/M_{\odot}) > 10.7$) subsamples, and plot the AGN and SFG fractions for the two mass bins in Fig. \ref{fig5}.
It can be seen that SFG fraction in low-mass galaxies is much higher at all radii than that in massive galaxies, and their trends along with the projected radius are completely different. The SFG fraction in lower mass galaxies tends to be higher at larger radius, whereas no significant change in SFG fraction is found in massive galaxies. Additionally, the mass dependence in AGN fraction is also confirmed in Fig. \ref{fig5}. In detail, a systematically lower AGN fraction ($\sim 10\%$) is found at all radii for less massive group galaxies, and it does not vary with projected radius. For the massive group galaxies, the AGN fraction has doubled, with a weak rising trend with group-centric radius. Additionally, the fraction of early-type (`UnClass') galaxies in high-mass subsample becomes higher, with a weaker concentration trend.  The fraction of composite galaxies show no obvious variation with the normalized projected radius, with a constant fraction of $\sim 5\%$ in low and high mass subsamples. Since the number of composite galaxies is nearly half of that of AGNs, particularly for the poor groups with $3 \leqslant N_{\rm rich} \leqslant 4$ (see Table \ref{table1}), the variations of AGN fraction with group richness and $R/R_{200}$ will become more uncertain when the composite population exhibits a different varying trend from the AGN sample. Fortunately, it can be found that the variations of composite population with $N_{\rm rich}$ and $R/R_{200}$ obey a similar trend with the AGN sample (see Figure 2, 5, 6, and 7).

For local clusters of galaxies, \citet{Hwang12} checked the morphological dependence of radial $f_{\rm AGN}$ distribution with the SDSS DR7 samples, and found that AGN fraction in early-type galaxies is much lower than that in late-type galaxies within one virial radius. To see the possible morphological dependence in the environment of groups, we also split our sample of group galaxies in the $N_{\rm rich} \geqslant 5$ groups into early-type (E $+$ S0) and late-type (Sab $+$ Scd) galaxies, and reestimate the fractions of different classes. It is found in Fig. \ref{fig6} that AGN fraction is almost the same for different morphological subsamples, and so does the fraction of composite galaxies. Not surprisingly, SFG fraction in late-type galaxies is much higher than that in early-type galaxies, and the late-type SFGs are more likely to be located in outskirts of groups.

\subsection{Distributions of environmental density }

\begin{figure}
\center
\includegraphics[width=7.5cm]{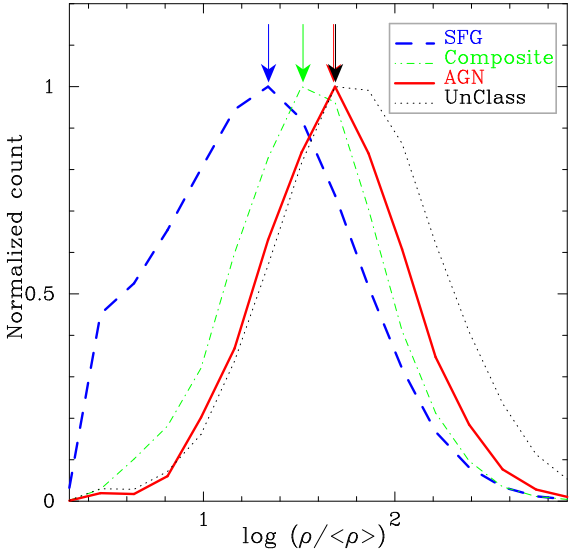}
\caption{The normalized number of galaxies as a function of the normalized environmental density of the galaxies. The smoothing scale is 1.5 Mpc.}
\label{fig7}
\end{figure}

The distributions of environmental density for the different classes of group galaxies can shed light on the fueling mechanisms of AGN and star formation activities.
Different smoothing scales adopted in calculating the density field represent different environments (see Eq.(4), (7) and Fig. 7 in \citealt{Tempel12}). Smaller smoothing lengths represent the group scales, and larger one correspond to the cluster and/or supercluster environments. \citet{Tempel17} estimated the environmental luminosity density field using the SDSS $r$-band luminosities adopting the smoothing lengths of 1.5, 3, 6, and 10 Mpc. We adopt the smallest smoothing scale of 1.5 Mpc to exhibit the environments of galaxies in different classes. Fig. \ref{fig7} shows the distributions of normalized densities for different classes of galaxies. The arrows indicate the peaks of the density distributions for different galaxy classes. It can be seen in Fig. \ref{fig7} that the group AGNs are likely to reside in the regions with higher densities, whereas the SFGs are systematically distributed in the regions of lower density.
This conclusion is consistent with the previous result shown in Fig. \ref{fig3} that the AGNs in groups are likely to be located in inner regions.
Although the AGNs and the early-type galaxies in `UnClass' population peak at similar density, the early-type galaxies are found to have a wider coverage of density, and they are likely to reside in the regions with higher densities.

\subsection{Star formation properties as functions of richness} \label{prop}

Based on the spectral analyses, the star formation properties, such as age and mass of stellar populations, star formation rate (SFR), and specific SFR (i.e. the SFR normalized to stellar mass, sSFR hereafter), can be derived for each group member galaxy. In this section, we analyze how the mean values of stellar mass ($M_{*}$), SFR, sSFR, and the strength of the continuum break at 4000 \AA~, $D_{n}(4000)$, vary with group richness. As mentioned above, the continuum break at 4000 \AA~ is a prominent feature in galaxy spectra, which can be treated as a powerful estimator of stellar age because the 4000 \AA~ break appears very weak for young stellar populations and strong for old metal-rich galaxies. \citet{Kauf03c} used the amplitude of the 4000 \AA~ break and the strength of the ${\rm H}\delta$ absorption line as diagnostics of the stellar populations within a galaxy. The SDSS $z$-band is the least sensitive to the effects of dust attenuation, \citet{Kauf03c} first used the measured $D_{n}(4000)$ and ${\rm H}\delta_{\rm A}$ indices to obtain a maximum-likelihood estimate of the $z$-band mass-to-light ratio ($M/L$) for each galaxy, then derived the stellar mass ($M_{*}$) by combining with the $z$-band absolute magnitude and dust attenuation $A_{z}$. For the AGN host galaxies, to avoid line emission from central AGNs, \citet{Brin04}  used the measured $D_{n}(4000)$ value to estimate their SFRs. They constructed a relationship between SFR/$M_{*}$ and $D_{n}(4000)$ to estimate the sSFR, thus total SFR for the AGN host galaxy can be yielded.  Considering that both $D_{n}(4000)$ and ${\rm H}\delta$ absorption strength have been corrected for the observed contributions of the emission lines in their bandpasses, the contamination of central AGN emission can be negligible in the measurements of stellar mass and SFR.


\begin{figure}
\center
\includegraphics[width=9cm]{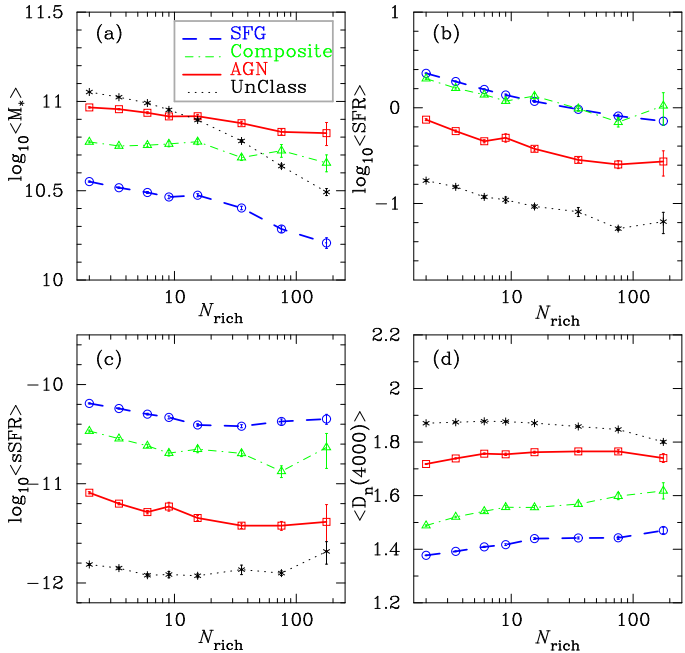}
\caption{Distributions of stellar mass $\langle M_{*} \rangle$, SFR, sSFR, and stellar age indicator $D_{n}(4000)$,  as a function of group richness. The error bars correspond to standard uncertainties of the mean values of these four quantities for each richness bin.}
\label{fig8}
\end{figure}

Panel (a) in Fig. \ref{fig8} gives the mean values of stellar mass for all classes of galaxies as a function of group richness. Our group sample has been divided into 8 richness bins (see Table \ref{table1}), and the error bar represents standard uncertainty of the derived mean value for each richness bin. In general, the mean stellar mass, $\langle M_{*} \rangle$, for all AGN host galaxies are $8.05 \times 10^{10}$ M$_{\odot}$, about 2 times larger than that for all SFGs in groups, ~$2.74 \times 10^{10}$ M$_{\odot}$.  The mean mass for the composite galaxies is lower than that of the AGNs. For poor groups ($N_{\rm rich} \leqslant 10$), the `UnClass' early-type galaxies are found to have higher mean stellar mass. This result is consistent with the conclusion that more massive galaxies are more likely to host AGNs than their lower mass counterparts \citep{Dunlop03,Kauf03b,Floyd04,Brusa09,Cardamone10,Xue10,Pimbblet13}. Four classes of galaxies show a decreasing trend of the mean stellar mass along with richness. In particular,  the SFGs exhibit a very tight anti-correlation between mean stellar mass and group richness, corresponding to a linear fitting of ${\rm log}\langle M_{*} \rangle = (-0.181 \pm 0.019) {\rm log}N_{\rm rich} + (10.641 \pm 0.025)$.  According to Fig. \ref{fig4}, a larger fraction of SFGs is presented at larger $R/R_{200}$, and the majority of these SFGs are found to be late-type galaxies with lower stellar masses (see also Fig. \ref{fig5}a and Fig. \ref{fig6}b). For the AGN host galaxies in groups, a similar declining trend of mean stellar mass can also be found, and its linear fitting can be described as: ${\rm log}\langle M_{*} \rangle = (-0.084 \pm 0.006) {\rm log}N_{\rm rich} + (11.00 \pm 0.008)$.

Fig. \ref{fig8}(b) exhibits the mean SFRs of different classes of galaxies as a function of group richness. It is clear that the mean SFRs of the SFGs are much higher than those of the AGNs and `UnClass' early-type galaxies. For any specified richness of group, the AGN host galaxies in our spectroscopic sample are found to have intermediate $\langle SFR \rangle$ value between the SFGs and `UnClass' galaxies. Compared with the SFGs, the star formation activities in AGNs are systematically lower. The mean SFR value for the SFGs is about 3 times of that for the AGNs. The mean SFR of the SFGs shows a clear decline with group richness, with a linear fitting of ${\rm log}\langle \rm SFR \rangle = (-0.258 \pm 0.016) N_{\rm rich} + (0.403 \pm 0.021)$. For the AGN host galaxies, the mean SFRs have similar trend with group richness, with a linear fitting of ${\rm log}\langle \rm SFR\rangle = (-0.235 \pm 0.033) N_{\rm rich} - (0.117 \pm 0.044)$. The morphological distributions for Unclass, AGN, and SFG populations can well explain the fact that the AGN host galaxies have an intermediate mean SFRs between the Unclass (elliptical) galaxies and  SFGs.  As shown in the weighted probability distributions (Fig. \ref{fig2}), the AGN population contains $\sim$ 54\% late-type (Sab $+$ Scd) galaxies, whereas majority ($\sim$ 79\%) of the SFGs are late-type galaxies.

It is well known that the SFRs in SFGs correlate tightly with their stellar masses.
\citet{Schreiber15} unveiled a universal SFR$-$M$_*$ relation, with a flattening of the main sequence at high masses in the local universe, therefore the specific star formation rates (sSFRs) for local massive ($\log M_*/M_{\odot} > 10.5$) SFGs tend to lower. Fig. \ref{fig8}(c) shows the mean sSFRs of four classes of galaxies as a function of group richness. The mean sSFRs in each richness bin is calculated by $\langle \rm sSFR \rangle = \langle \rm SFR \rangle/\langle M_{*} \rangle$. Compared with the SFGs, average sSFR of the AGN host galaxies have been reduced about one order of magnitude. The `UnClass' galaxies are proved to be completely inactive at all richness. For both SFGs and AGN host galaxies, their average sSFRs are found to be higher in poor groups than in rich ones. This declining trend might imply that a denser gravity environment is likely to be against star formation.

Fig. \ref{fig8}(d) displays the mean $D_{n}(4000)$, as a function of the group richness.
Taking $D_{n}(4000)$ as an indicator of average stellar age, the quiescent `UnClass' galaxies possess the oldest stellar population because their star formation activities have been completed long time ago. The AGN host galaxies are found to have a constant mean value of $D_n(4000)$ at all richness, indicating that the star formation histories for the AGN host galaxies seem to be insensitive to large scale environment of gravity. On the other hand, the SFGs in groups are found to have younger stellar population, exhibiting a weak rising trend of $D_{n}(4000)$ with group richness.

\section{DISCUSSION}

We have presented the incidences, morphologies, and star formation properties for different
categories of group galaxies as functions of the group richness. The basic assumption is that the group richness is a direct observational quantity that can represent the gravity environment on a scale of galaxy group. The virial mass and dispersion of radial velocity are commonly adopted to describe group/cluster environment of gravity, but they are model-dependent, not directly observable quantities. They can be derived if we assuming that the groups are virilized systems.

\begin{figure}
\center
\includegraphics[width=7cm]{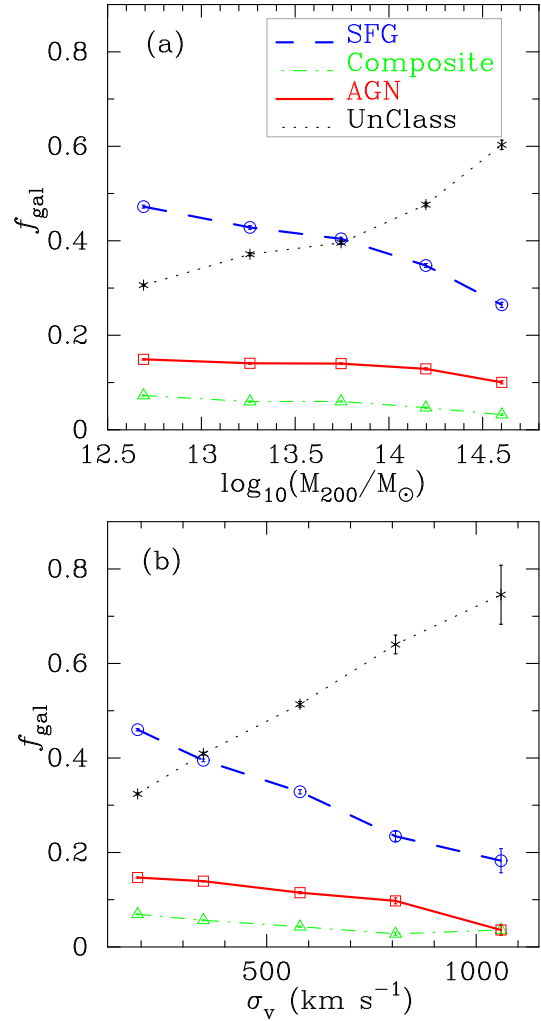}
\caption{The fractions of different classes of galaxies, $f_{\rm gal}$, as functions of group mass ($M_{200}$) and group radial velocity deviation ($\sigma_{v}$). Blue circles (dashed line), green triangles (dot-dashed line), red rectangles (solid line), and black crosses (dotted line) denote the SFGs, Composites, AGNs, and `UnClass' galaxies, respectively.}
\label{fig9}
\end{figure}

\citet{Tempel17} also estimated the virial mass within $R_{200}$, $M_{200}$, for each group, assuming an NFW mass density profile \citep{NFW97}. The velocity dispersion can be described by the measured rms deviation of radial velocity \citep{Tempel12},  assuming an isotropic dynamics in groups. To demonstrate the reasonability of using group richness as the indicator of gravity environment, we present the fractions of different classes of galaxies in Fig. \ref{fig9}, as functions of virial mass ($M_{200}$) and radial velocity deviation ($\sigma_v$).
We split our group sample into five bins according to group virial mass, $\log(M_{200}/M_{\odot})$:  (1) $<$13.0, (2) [13.0,13.5), (3) [13.5, 14.0), (4) [14.0, 14.5), and (5) $\geqslant 14.5$.
Compared with Fig. \ref{fig1}, Fig. \ref{fig9}(a) shows very similar trends of population fractions, suggesting that the group richness is a good tracer of group mass.
It is confirmed that the fractions of AGN host galaxies decline slightly from $15\%$ to $12\%$ within a wide range of group mass, $12.5 \leqslant \log(M_{200}/M_{\odot}) \leqslant 14.5$, and a significant decline in AGN fraction for poor clusters ($\log(M_{200}/M_{\odot}) > 14.5$) is also presented. Compared with the trend of population fraction along with richness (see Fig.2), a weaker decline at larger $M_{200}/M_{\odot}$ can be found in Fig. 10(a). According to mass-richness relation shown in Fig. 1, the highest mass bin ($\log M/M_{\odot} \geqslant 14.5$) in Fig. 10(a) corresponds to two richness bins (i.e., $51 \leqslant N_{\rm rich} \leqslant 100$ and $101 \leqslant N_{\rm rich} \leqslant 254$). The mean galaxy fraction in the poor clusters with $\log M/M_{\odot}> 14.5$ are equivalent to the smoothed value of galaxy fraction at two highest richness bins. The trends of galaxy fraction along with richness and group mass are self-consistent. Our $f_{\rm AGN}$ trend is in accordance with the recent result by \citet{Gordon18} (see Fig. 4 therein) who found that lower mass groups, in general, have higher AGN fraction than higher mass groups. It should be noted that a certain number of low S/N AGNs (e.g. the Low Ionization Nuclear Emission Regions, LINERs) are included in our AGN sample, and our average AGN fraction is about two times more than that in \citet{Gordon18}.


Meanwhile, our group sample can be divided into five bins according to the radial velocity dispersion, $\sigma_v/(\rm km~s^{-1})$: (1) $<250$, (2) [250, 500), (3) [500, 750), (4) [750,1000), and $\geqslant 1000$. The incidence of each galaxy class is plotted in Fig. \ref{fig9}(b), and a similar declining trend varying with velocity dispersion is found for both SFGs and AGNs. A slight decline of $f_{\rm AGN}$ is confirmed at $\sigma_v \leqslant 800$  km~s$^{-1}$, as well as a strong decline at $\sigma_v \geqslant 1000$ km~s$^{-1}$.
Based on the SDSS data, \citet{Popesso06} detected a decrease in $f_{\rm AGN}$ within a velocity dispersion range from $\sim 200$ to 600 km~s$^{-1}$, and a constant $f_{\rm AGN}$  for higher velocity dispersion. This result has been interpreted as a consequence of the galaxy-galaxy merger inefficiency in rich clusters.
Our results confirm this decreasing trend at low velocity dispersion, and extend this trend to $\sigma_v > 1000$ km~s$^{-1}$. \citet{Martini07} also found that some clusters with lower velocity dispersions have higher AGN fractions. Using the Northern Sky Optical Cluster Survey (NoSOCS), \citet{Lopes17} verified the dependence of the AGN fraction on cluster velocity dispersion, finding a constant  $f_{\rm AGN}$  at $ \sigma \leqslant 650 - 700 {\rm km s}^{-1} $ and a strong decline for higher mass clusters with $\sigma > 700 {\rm km s}^{-1}$. This decline at high velocity dispersion is confirmed by this work.
However, some previous studies found that the frequency of low activity AGNs does not correlate with environment \citep[e.g.][]{Shen07,Miller03}. Because these early investigations are based on small samples of AGNs, the larger relative errors in $f_{\rm AGN}$ prevent the discovery of the sophisticated trend that $f_{\rm AGN}$ slightly declines with group environnment.

The variations of $f_{\rm AGN}$ in groups with richness, virial mass, and velocity dispersion are proved to be highly self-consistent, which has been well shown in Fig. \ref{fig1} and Fig. \ref{fig9}. This suggests that the richness ($N_{\rm rich}$) can be treated as an observable indicator of gravity environment of galaxy group, equivalent to virial mass of group.

\section{CONCLUSION}

Using a large sample of SDSS galaxy groups at $z<0.2$, which is constructed from the \citet{Tempel17} group catalog and the MPA/JHU database of the SDSS spectroscopic galaxy sample, we investigate the spatial distributions, morphologies, and star formation properties of the AGN host galaxies in groups. All member galaxies in groups are spectroscopically classified into four populations: AGN host galaxies, Composites, SFGs, and `UnClass' galaxies.
The group richness is a direct observable tracer of gravity environment on a scale of group. The incidences, morphological percentage, environmental densities, and star formation properties (e.g. stellar mass, SFR, sSFR, and stellar age) are investigated as functions of group richness for these four classes of galaxies, particularly for the AGN host galaxies and SFGs. Our main conclusions are summarized as follows:

(1) The AGN fractions are roughly constant ($\sim 14.5\%$) for poor groups with $3 \leqslant N_{\rm rich} \leqslant 10$, and decline slightly with richness for the groups and poor clusters with $3 \leqslant N_{\rm rich} \leqslant 100$. A significant drop of AGN fraction is found for the clusters with $N_{\rm rich} > 100$. The weak decreasing trend along with group richness supports the point that galaxies in poor groups retain larger reservoirs of cold gas to fuel AGN activity than their counterparts in rich groups/clusters.
The mean value of SFG fraction is found to be about 2 times larger than that of AGN fraction, and it decreases significantly with richness, which can be interpreted by the some popular mechanisms of environmental quenching, such as harassment and ram pressure stripping.

(2) Majority of AGNs are found to reside in spheroidal and bulge-dominated disc galaxies (S0 $+$ Sab), whereas $\sim 80\%$ SFGs are found to be late-type disc galaxies (Sab $+$ Scd).

(3) Distributions of normalized group-centric radius show that the AGNs in poor groups with $5 \leqslant N_{\rm rich} \leqslant 10$ are likely to locate in inner regions, which is different with the SFGs in poor groups. The AGNs in rich groups distribute slightly away from the center. The AGN fraction does not vary with the projected radius, whereas the SFG fraction tends to be higher at larger radius. No significant change in $f_{\rm AGN}$ is found for the groups with different richness.

(4) A significant mass dependence in radial profiles of the AGN and SFG fractions are found. The SFG fraction in lower mass galaxies is much higher at all radii than that in massive galaxies, and SFG fraction in late-type galaxies with lower masses exhibits a clear rising trend with group-centric radius. The AGN fraction has doubled for the massive group galaxies, and tends to be slightly higher at larger radius.

(5) The distribution of environmental densities shows that the group AGNs are likely to reside in a denser environment relative to the SFGs.

(6) In general, the AGN host galaxies have a larger average stellar mass than the SFGs, and have intermediate mean SFRs between the SFGs and `UnClass' early-type galaxies. The mean stellar ages in AGN host galaxies are found to be unchangeable at all richness, indicating that the star formation histories with the AGN host galaxies are insensitive to large scale environment of gravity. The SFGs exhibit a weak rising trend of $D_{n}(4000)$ with group richness.

\section{ACKNOWLEDGMENTS}

We would like to thank the anonymous referee for his/her very thorough reading and valuable suggestions which help us to improve this paper a lot.
We are very grateful to Prof. Xu Zhou at the National Astronomical Observatories of the Chinese Academy of Sciences (NAOC) and Prof. Xu Kong at University of Science and Technology of China (USTC) for their enlightening discussion and instructive suggestions. This work has been supported by the National Natural Science Foundations of China (NSFC) (Nos. 11173016, 11873032, 11433005, 11373024 and 11233003) and by the Special Research Foundation for the Doctoral Program of Higher Education (grant No. 20133207110006).

This research has made use of the SAO/NASA Astrophysics Data System. Funding for the SDSS and SDSS-III was provided by the Alfred P. Sloan Foundation, the Participating Institutions, the National Science Foundation, the U.S. Department of Energy, the National Aeronautics and Space Administration, the Japanese Monbukagakusho, the Max Planck Society, and the Higher Education Funding Council for England. A list of participating institutions can be obtained from the SDSS Web Site http://www.sdss.org/.

\end{document}